\documentclass[twocolumn,10pt,aps,preprintnumbers,showpacs,pre,floatfix,amsmath,amssymb,notitlepage,superscriptaddress]{revtex4-1}
\usepackage{epsfig}
\usepackage{subfigure}
\usepackage{amsmath}
\usepackage{color}
\usepackage{amssymb}
\usepackage{setspace}
\usepackage{graphicx}
\usepackage{dcolumn}
\usepackage{bm}
\usepackage{times}
\usepackage{enumerate}
\usepackage{footnote}
\usepackage{cancel}
\input epsf

\newcommand{\appropto}{\mathrel{\vcenter{
  \offinterlineskip\halign{\hfil$##$\cr
    \propto\cr\noalign{\kern2pt}\sim\cr\noalign{\kern-2pt}}}}}

\graphicspath{{figures/}}

\begin{document}

\author{Per Sebastian Skardal}
\email{persebastian.skardal@trincoll.edu} 
\affiliation{Department of Mathematics, Trinity College, Hartford, CT 06106, USA}

\author{Dane Taylor}
\affiliation{Carolina Center for Interdisciplinary Applied Mathematics, Department of Mathematics, University of North Carolina, Chapel Hill, NC 27599, USA}

\author{Jie Sun}
\affiliation{Department of Mathematics, Clarkson University, Potsdam, NY 13699, USA}
\affiliation{Department of Physics, Clarkson University, Potsdam, NY 13699, USA}

\author{Alex Arenas}
\affiliation{Departament d'Enginyeria Inform\`{a}tica i Matem\`{a}tiques, Universitat Rovira i Virgili, 43007 Tarragona, Spain}

\title{Collective frequency variation in network synchronization and reverse PageRank}

\begin{abstract}
A wide range of natural and engineered phenomena rely on large networks of interacting units to reach a dynamical consensus state where the system collectively operates. Here we study the dynamics of self-organizing systems and show that for generic directed networks the collective frequency of the ensemble is {\it not} the same as the mean of the individuals' natural frequencies. Specifically, we show that the collective frequency equals a weighted average of the natural frequencies, where the weights are given by an out-flow centrality measure that is equivalent to a reverse PageRank centrality. Our findings uncover an intricate dependence of the collective frequency on both the structural directedness and dynamical heterogeneity of the network, and also reveal an unexplored connection between synchronization and PageRank, which opens the possibility of applying PageRank optimization to synchronization. Finally, we demonstrate the presence of collective frequency variation in real-world networks by considering the UK and Scandinavian power grids.
\end{abstract}


\maketitle

\section{Introduction}\label{sec1}

The emergence of synchronization in ensembles of dynamical units is a universal phenomenon that is vital to the functionality of many natural and man-made systems~\cite{Strogatz2003,Pikovsky2003,Arenas2008PR}. In addition to the ability of the individuals that make up such systems to operate in unison, in many instances the particular frequency or velocity with which they evolve is crucial. For example, the sources and loads that make up power grids must reach consensus to avoid power failures, but reaching a common frequency alone is not enough; the system is most efficient near a certain reference frequency of approximately $50$ - $60$ Hz, and may fail if the collective dynamics are too far from this range~\cite{Rohden2012PRL,Motter2013NaturePhysics}. In a wide variety of disciplines, from biology and neuroscience to mechanical and electrical engineering, there are vital systems whose functionality is jeopardized if the collective frequency or velocity differs too much from a given reference frequency; examples include brain dynamics, cardiac excitation, consensus networks, and coordination of muscle movements in the digestive track~\cite{Antzoulatos2014Neuron,Karma2007PT,OlaftiSaber007IEEE,Aliev200JTB}. In the case of cardiac excitation, for instance, rapid oscilations can give rise to dynamical instabilities that often precede ventricular fibrillation and eventually heart failure. 

In the majority of works studying the dynamics of network synchronization, it is often assumed that the collective frequency of the synchronized state is precisely the mean natural frequency of the individual units~\cite{Pikovsky2003,Arenas2008PR,Kuramoto1984}. In other words, the synchronized state reaches an oscillation frequency that is equal to the unweighted average of the oscillation frequencies of the individual elements when acting in isolation, i.e., uncoupled. In this Article we study the collective frequency of self-organizing systems of oscillators and show that it is {\it not} in general equal to the mean of the individuals' natural frequencies. We find that collective frequency variation is a consequence of the directedness of network and heterogeneity of the dynamics. For networks lacking either, e.g., undirected networks or identical oscillators, we find that that the collective frequency does recover the mean oscillator frequency~\cite{Ott2008Chaos,Restrepo2005PRE}. Importantly, systems with directed connections and non-identical agents are ubiquitous~\cite{Strogatz2003}, and therefore collective frequency variation is a fundamental--yet unexplored--property of real-world self-organizing systems~\cite{Restrepo2006Chaos,Assenza2011SciRep}. 

To investigate this phenomenon, we consider the general linearized dynamics of $N$ coupled units, $x_i$, for $i=1,\dots,N$, given by
\begin{align}
\dot{x}_i=\omega_i-K\sum_{j=1}^NL_{ij}x_j,\label{eq:01}
\end{align}
where $\omega_i$ is the natural frequency of oscillator $i$, $K$ is the global coupling strength, and $L$ is the network Laplacian matrix. The entries of $L$ are defined $L_{ij}=\delta_{ij}k_{i}^{\text{in}}-A_{ij}$, where $A_{ij}$ is the network adjacency matrix and $k_i^{\text{in}}=\sum_{j=1}^NA_{ij}$ is the in-degree of node $i$. We also define the out-degree of node~$i$, $k_i^\text{out}=\sum_{j=1}^NA_{ji}$. We assume the network encoded by $A$ to be strongly-connected~\cite{Newman2003SIAM}. In principle, our analysis allows the network to be directed and weighted, although unless otherwise noted we will focus on the case of unweighted edges: $A_{ij}=1$ if a directed link $j\to i$ exists, and otherwise $A_{ij}=0$. We note that there are several ways to define a Laplacian matrix for directed networks~\cite{Chung2005}; we study a version that is appropriate for the dynamics of interest. These linearized dynamics represent a versatile description of a wide range of dynamical processes on networks~\cite{Grabow2012PRL,Grabow2015PRE}. For instance, Eq.~(\ref{eq:01}) can be obtained from linearizing self-organizing systems around the synchronized manifold, for instance the Kuramoto model which serves as a model for a wide range of synchronization phenomena including power grid dynamics~\cite{Dorfler2013PNAS,Skardal2015SA}, as well as other systems with more general coupling which are utilized in modeling excitable- and reaction-diffusion-type systems~\cite{Kopell2002,Skardal2014PRL,Skardal2015PRE}. This linear relaxation has been found to accurately capture the dynamics of the system, provided that initial conditions are within the basin of attraction of the synchronized state~\cite{Menck2013NatPhys}. In the case of a network of coupled oscillators, this tends to be particularly robust, capturing the dynamics provided that the overall coupling is not too small in comparison to the spread in the natural frequencies (which we illustrate in Sec.~\ref{sec5}).
 
We study the frequency-synchronized state, given by $\dot{x}_1=\dots=\dot{x}_N$, and quantify the collective frequency variation by examining $\Omega-\langle\omega\rangle$, where $\Omega$ denotes the collective frequency of the synchronized population and $\langle\omega\rangle=N^{-1}\sum_i\omega_i$ is the mean natural frequency. We call this difference the {\it collective frequency variation}. We show that under generic conditions which are present in most practical application, when the frequencies $\omega_i$ are non-identical and the in- and out-degrees $k_i^\text{in}$ and $k_i^\text{out}$ are not perfectly balanced, then $\Omega-\langle\omega\rangle\ne0$. However, when the in- and out-degrees match for each node in the network, $k_i^{\text{in}}=k_i^{\text{out}}$, then the collective frequency variation vanishes, i.e., $\Omega=\langle\omega\rangle$, for any choice of frequencies. We calculate the collective frequency variation directly from Eq.~(\ref{eq:01}) and show that $\Omega-\langle\omega\rangle$ is given by a weighted average of the natural frequency vector, where the weights correspond to entries of the first left singular vector $\bm{u}^1$ of $L$ that is associated with the trivial singular value $\sigma_1=0$. We find that $\bm{u}^1$ represents an out-flow centrality measure, and in fact the entries of $\bm{u}^1$ are often well-approximated by the out-to-in-degree ratio, $u_i\appropto k_i^\text{out}/k_i^\text{in}$. Interestingly, the first-left-singular-vector centrality is a reverse analogue of Google's PageRank centrality~\cite{Brin1998}, which provides a cornerstone to Google's ranking of webpages and favors nodes with strong in-flow~\cite{Gleich2015SIAM}. These findings reveal an interesting and surprising link between synchronization dynamics and PageRank, paving a path for new theoretical exploration and the possibility of applying well-established PageRank methods to synchronization. We will also demonstrate the presence of collective frequency variation in real-world UK and Scandinavian power grid networks. However we emphasize that our findings fit in a much broader and more interdisciplinary framework.

The remainder of this Article is organized as follows. In Sec.~\ref{sec2} we derive the collective frequency variation of a network. In Sec.~\ref{sec3} we study the range of possible collective frequency variation for a given network structure. In Sec.~\ref{sec4} we show that the weights that contribute to the collective frequency admit a centrality that is the reverse analogue of Google's PageRank centrality. In Sec.~\ref{sec5} we study collective frequency variation in the power grid as a real example. In Sec.~\ref{sec6} we conclude with a discussion of our results.

\section{Derivation of collective frequency variation}\label{sec2}

We begin by writing Eq.~(\ref{eq:01}) in vector form,
\begin{align}
\dot{\bm{x}}=\bm{\omega}-KL\bm{x}.\label{eq:02}
\end{align}
Our aim is to calculate the collective frequency of the synchronized population, and therefore we propose the ansatz
\begin{align}
\bm{x}(t) = \bm{x}^*+\Omega\bm{1}t,\label{eq:03}
\end{align}
where $\bm{x}^*$ is a vector encoding the steady-state value of each $x_i$ in an appropriate rotating frame, $\bm{1}=[1,\dots,1]^T$, and $\Omega$ is the collective frequency. To proceed, we will utilize the pseudoinverse $L^\dagger$ of the Laplacian matrix, which satisfies $LL^\dagger L=L$ and $L^\dagger LL^\dagger=L^\dagger$~\cite{BenIsrael1974}. In the undirected case, $L^\dagger$ can be found using the eigenvalue decomposition of $L$, whereas in the more general case of a directed network, $L^\dagger$ is formulated in terms of the singular value decomposition (SVD) of $L$. In particular, if $L=U\Sigma V^T=\sum_{j=2}^N\sigma_j\bm{u}^j\bm{v}^{jT}$, where $\sigma_j\ge0$ are the singular values which are ordered $0=\sigma_1<\sigma_2\le\dots\le\sigma_N$ and make up the diagonal entries of $\Sigma$, and $\bm{u}^j$ and $\bm{v}^{jT}$ are the corresponding left and right singular vectors that make up the columns of $U$ and $V$, respectively, then the pseudoinverse is given by $L^\dagger=V\Sigma^\dagger U^T=\sum_{j=2}^N\sigma_j^{-1}\bm{v}^j\bm{u}^{jT}$. An important distinction between $L$ and $L^\dagger$ is that, while $L$ maps all constant vectors to zero since its rows sum to zero, this is not generally true of $L^\dagger$, whose nullspace is nontrivial. Furthermore, the sets of singular vectors $\{\bm{u}^j\}_{j=1}^N$ and $\{\bm{v}^j\}_{j=1}^N$ (appropriately normalized) each form an orthonormal basis for $\mathbb{R}^N$.

\begin{figure*}[t]
\centering
\includegraphics[width=0.80\linewidth]{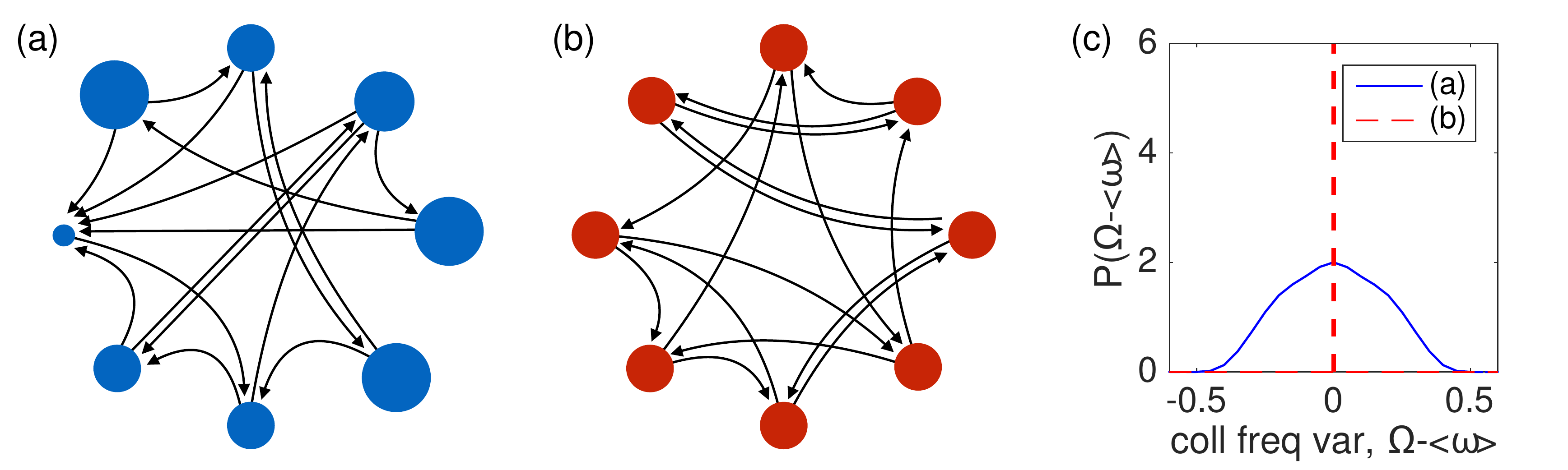}
\caption{(Color online) {\it Collective frequency variation.} (a),(b) Two networks of size $N=8$ with $16$ links. In (b), the in- and out-degrees match at each node, in particular $k_i^\text{in}=k_i^\text{out}=2$. In (a) this balance is broken, so $k_i^{\text{in}}\ne k_i^{\text{out}}$. Each node's area is proportional to the ratio $k_i^\text{out}/k_i^\text{in}$, which represents a mean field approximation to the first left singular vector $\bm u^1$ of the Laplacian matrix $L$. (c) The density $P(\Omega)$ of collective frequencies $\Omega$ observed in networks (a) and (b) (solid blue and dashed red, respectively) for different permutations of a normally distributed frequency vector $\bm \omega$ with mean $\langle\omega\rangle=0$ and variance $\sigma^2=1$. We find $\Omega$ to relate closely to the alignment of $\bm \omega$ with vector $\bm u^1$, which represents an out-flow centrality.}
\label{fig1}
\end{figure*}

Proceeding with the analysis, we insert Eq.~(\ref{eq:03}) into Eq.~(\ref{eq:02}) and rearrange to obtain
\begin{align}
\bm{\omega}-\Omega\bm{1}=KL\bm{x}^*.\label{eq:04}
\end{align}
Left-multiplying by $LL^\dagger$, and using that $LL^\dagger L=L$, we find
\begin{align}
LL^\dagger\left(\bm{\omega}-\Omega\bm{1}\right)=KL\bm{x}^*.\label{eq:05}
\end{align}
Equations~(\ref{eq:04}) and (\ref{eq:05}) thus imply that
\begin{align}
(I-LL^\dagger)\bm{\omega}=\Omega(I-LL^\dagger)\bm{1}.\label{eq:06}
\end{align}
Next, since $\sigma_1=0$, the matrix $I-LL^\dagger$ can be simplified to $\bm{u}^1\bm{u}^{1T}$. Finally, we left-multiply Eq.~(\ref{eq:06}) by $\bm{1}$, rearrange, and subtract $\langle\omega\rangle$ from the right- and left-hand sides to obtain
\begin{align}
\Omega-\langle\omega\rangle=\frac{\langle\bm{u}^1,\bm{\omega}-\langle\omega\rangle\bm{1}\rangle}{\langle\bm{u}^1,\bm{1}\rangle},\label{eq:07}
\end{align}
where $\langle\bm{a},\bm{b}\rangle=\bm{a}^T\bm{y}=\sum_{i}a_ib_i$ denotes the inner product. This result is in good agreement with previous research on consensus systems. In particular, by differentiating Eq.~(\ref{eq:02}) with respect to time, using the initial condition $\dot{\bm{x}}(0) = \omega-KL\bm{x}(0)$, and noting that the first left singular vector and first left eigenvector are equal, we find that our derivation of Eq.~(\ref{eq:07}) provides a complementary derivation of Eq.~(23) in Ref.~\cite{OlaftiSaber2004IEEE}.

Equation~(\ref{eq:07}) gives the collective frequency variation $\Omega-\langle\omega\rangle$ of a synchronized population as the projection of the natural frequency vector $\bm{\omega}-\langle\omega\rangle\bm{1}$ (shifted to have zero mean) onto the first left singular vector $\bm{u}^1$. The physical interpretation of Eq.~(\ref{eq:07}) is that the collective frequency variation is a weighted average of the natural frequencies, wherein the weights are proportional to the entries of $\bm{u}^1$. Thus, nodes with large entries in $\bm{u}^1$ contribute more to the collective frequency variation than those with small entries, allowing for non-zero values of $\Omega-\langle\omega\rangle$ provided that the entries of $\bm{u}^1$ are not identical. Furthermore, we can formulate the full range of collective frequencies for a given network as the maximum of $|\Omega-\langle\omega\rangle|$ over all choices of $\bm{\omega}$ with some fixed variance. As we will show below, the first left singular vector $\bm{u}^1$ induces a centrality measure for the network that is related to the out-flow of each node. Interestingly, we will show that this centrality is analogous to a ``reverse'' PageRank. In fact, it is equivalent to Google's PageRank centrality for the network obtained by reversing the direction of each link in the original network.

\section{Range of collective frequency variation}\label{sec3}

We demonstrate our main result, Eq.~\eqref{eq:07}, with a simple example using two small networks of size $N=8$, which are illustrated in Fig.~\ref{fig1}(a) and (b). Both networks contain $16$ links, yielding a mean in- and out-degree of $\langle k\rangle=2$; however, in network~(a) the links are made randomly so the in- and out-degrees at each node are not necessarily equal, while network (b) is balanced so that the links are made to satisfy $k_i^\text{in}=k_i^\text{out}=2$ for all $i$, but is still directed. For visual distinction, each node's area is proportional to the out-to-in-degree ratio $k_i^\text{out}/k_i^\text{in}$. Next, we draw a set of normally distributed natural frequencies with mean $\langle\omega\rangle=0$ and variance $\sigma^2=1$ and calculate for each network the collective frequency variation $\Omega-\langle\omega\rangle$ using Eq.~(\ref{eq:07}) for $10^4$ different permutations of these frequencies. In Fig.~\ref{fig1}(c), we plot the observed density $P(\Omega-\langle\omega\rangle)$ for networks (a) and (b) (solid blue and dashed red, respectively). In the generic case, network (a), where in- and out-degrees are not necessarily equal at each node, we observe a wide range of collective frequencies, while for network (b), where the balance $k_i^\text{in}=k_i^\text{out}$ is maintained, the collective frequency is zero in each case, resulting in a delta function $P(\Omega-\langle\omega\rangle)=\delta(\Omega-\langle\omega\rangle)$. This example highlights two important properties. First, the collective frequency variation is intimately linked with the directedness of a network: once the balance $k_i^\text{in}=k_i^\text{out}$ is broken, a non-zero value of $\Omega-\langle\omega\rangle$ should be expected. Second, the precise value of $\Omega-\langle\omega\rangle$ depends not only on the network and set of natural frequencies, but the arrangement of natural frequencies (dynamical heterogeneity) on the network. Therefore, for a fixed network and set of oscillator frequencies, depending on how the oscillators are assigned on the network, the system's collective frequency may either be faster or slower than the mean frequency.

A natural question to ask of a given network is: What is the range of possible collective frequency variations? We formalize this by considering for a given network, the magnitude of the maximum collective frequency variation across all frequency vectors with fixed variance $\sigma^2$, i.e., $\max_{\text{var}(\bm{\omega})=\sigma^2}|\Omega-\langle\omega\rangle|$. Inspecting Eq.~(\ref{eq:07}), it is straight-forward to see that the collective frequency variation is maximized when the shifted natural frequency vector $\bm{\omega}-\langle\omega\rangle\bm{1}$ is aligned with the first left singular vector $\bm{u}^1$. Thus, the choices of $\bm{\omega}$ that maximize $|\Omega-\langle\omega\rangle|$ with mean $\langle\omega\rangle$ and variance $\sigma^2$ are precisely 
\begin{align}
\bm{\omega}_\text{max}=\pm\sqrt{N}\sigma\frac{\bm{u}^1-\langle u^1\rangle\bm{1}}{\|\bm{u}^1-\langle u^1\rangle\bm{1}\|}+\langle\omega\rangle\bm{1},\label{eq:08}
\end{align}
where $\langle u^1\rangle=N^{-1}\sum_iu_i^1$ and the + and - symbols correspond to maximizing and minimizing $\Omega-\langle\omega\rangle$, respectively (that is, assuming $u_i^1>0$ for each $i$). This yields a collective frequency variation range of
\begin{align}
\max_{\text{var}(\bm{\omega})=\sigma^2}|\Omega-\langle\omega\rangle|=\sigma\sqrt{1-N\langle u^1\rangle^2}\Big/\sqrt{N\langle u^1\rangle^2}.\label{eq:09}
\end{align}

\begin{figure}[t]
\centering
\includegraphics[width=0.49\linewidth]{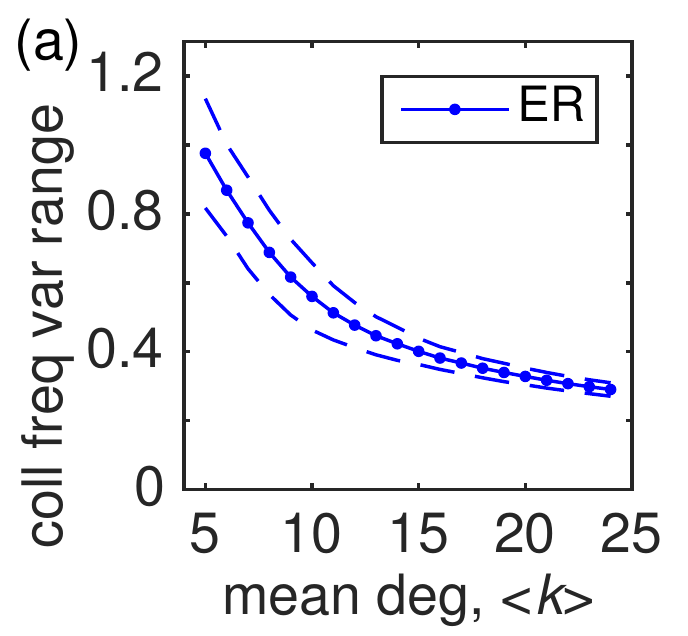}
\includegraphics[width=0.49\linewidth]{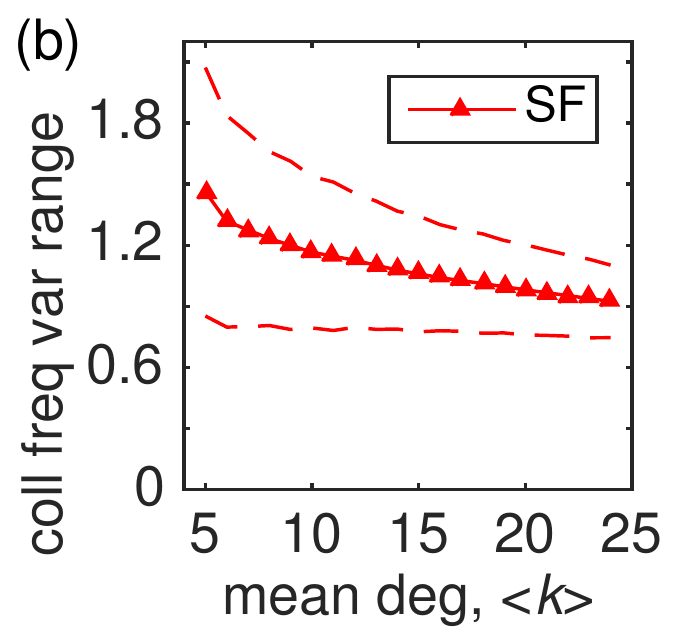}
\caption{(Color online) {\it Range of ollective frequency variation.} For (a) ER and (b) SF networks of size $N=200$ and various mean degrees, the collective frequency variation range $\max_{\text{var}(\bm{\omega})=\sigma^2}|\Omega-\langle\omega\rangle|$ for $\sigma^2=1$.} \label{fig2}
\end{figure}

To investigate how the range of collective frequency variation depends on network structure, we consider a variety of Erd\H{o}s-R\'{e}nyi~\cite{Erdos1960} (ER) and scale-free (SF) networks. ER networks are constructed using a link probability $p$ that describes the probability of directed link $j\to i$ existing. SF networks are built using the configuration model~\cite{Molloy1995} for target in- and out- degrees drawn from the distribution $P(k)\propto k^{-\gamma}$ for $k\ge k_0$, where $k_0$ is an enforced minimum degree. The mean degree for ER and SF networks can be tuned according to $\langle k\rangle=(N-1)p$ and $\langle k\rangle=(\gamma-1)k_0/(\gamma-2)$, respectively. In our experiment, we fix $\gamma=3$ and construct networks of size $N=200$ with various mean degrees and compute the collective frequency range according to Eq.~\eqref{eq:09} with $\sigma^2=1$. In Fig.~\ref{fig2}(a) and (b), we plot the results for over $1000$ ER and SF network realizations, respectively; we denote the mean and standard deviations using the symbols and dashed curves, respectively. For both network families, the collective frequency variation range tends to increase as the networks become more sparse. The central difference we observe is that both the mean collective frequency variation range and its standard deviation tend to be larger for SF networks than for ER networks. This suggests that structural heterogeneity has an amplifying effect on the range of collective frequency variation for a network -- however this effect can be mitigated on average by saturating the network structure: as the average connectivity increases, the range of collective frequency variation diminishes.

To better understand the role of network structure in determining collective frequency variation, we ask the following: For which network structures is the collective frequency variation exactly zero? That is, which network structures yield $\Omega-\langle\omega\rangle=0$ regardless of the choice of $\bm{\omega}$? From Eq.~(\ref{eq:07}), it follows that $\Omega-\langle\omega\rangle=0$ for any $\bm{\omega}$ whenever the entries of $\bm{u}^1$ are all identical, i.e., $\bm{u}^1\propto\bm{1}$. We note that since $L=D_\text{in}-A$, where $D_\text{in}=\text{diag}(k_1^\text{in},\dots,k_N^\text{in})$, and $\sigma_1=0$, then $\bm{u}^1$ must satisfy $\bm{u}^1=D_\text{in}^{-1}A^T\bm{u}^1$, or equivalently $\bm{u}^1$ is the leading right eigenvector of $D_\text{in}^{-1}A^T$. At each entry, we must have $u_i=\sum_{j=1}^NA_{ji}u_j/k_i^\text{in}$, and therefore by inserting $\bm{u}^1=c\bm{1}$ (for any $c\ne0$) it is easy to see then that $\bm{u}^1\propto\bm{1}$ implies that the network must be degree-balanced, i.e., $k_i^\text{in}=k_i^\text{out}$ for all $i$. The converse follows from a simple application of the Perron-Frobenius theorem~\cite{MacCluer2000SIAM}. Specifically, $\bm{u}^1\propto\bm{1}$ is a solution of the leading right eigenvalue equation for $D_\text{in}^{-1}A^T$, and the Perron-Frobenius theorem implies that it is in fact the unique solution, provided that the network is strongly connected. Therefore, any given network generically has zero collective frequency variation if and only if $k_i^\text{in}=k_i^\text{out}$ for all $i$.

\section{Singular vector centrality and Google's PageRank}\label{sec4}

Given the non-uniformity of each oscillator's contribution to a network's collective frequency variation, we now turn our attention to the properties of the first left singular vector $\bm{u}^1$, which dictates the contribution of each oscillator to the collective frequency variation. First, we note that the entries $u_i^1$ are positive, and thus $\bm{u}^1$ induces a centrality measure for the network. The positiveness of the entries follows from applying the Perron-Frobenious theorem~\cite{MacCluer2000SIAM} to the irreducible and non-negative matrix $D_{\text{in}}^{-1}A^{T}$ and noting that $\bm{u}^1$ is the leading right eigenvector of the matrix. The role of $\bm{u}^1$ as the leading right eigenvector of $D_\text{in}^{-1}A^T$ also elucidates its structural properties. In particular, Google's PageRank centrality -- which tends to favor nodes with strong in-flow -- is given by the leading right eigenvector $\bm v$ of the matrix $M=(q/N)\bm{1}\bm{1}^T+(1-q)D_\text{out}^{-1}A$, where $D_\text{out}=(k_1^\text{out},\dots,k_N^\text{out})$ and $q\in[0,1)$ is a damping factor~\cite{Gleich2015SIAM}. Formally the PageRank of a network represents the steady-state of a Markovian random-walk on the network. When the damping factor is set to zero and each directed link is reversed, the matrix $M$ from which PageRank is calculated is equal to $D_{\text{in}}^{-1}A^T$ (for which $\bm{u}^1$ is the leading right eigenvector). Thus, the centrality induced by the first left singular vector represents a reverse PageRank, i.e., the steady-state of a Markovian random walk on the network with each link reversed.
\begin{figure}[t]
\centering
\includegraphics[width=0.49\linewidth]{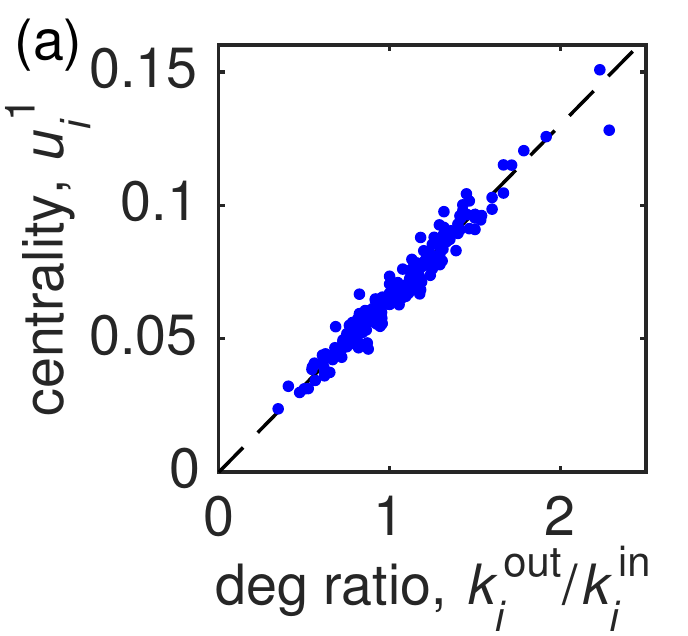}
\includegraphics[width=0.49\linewidth]{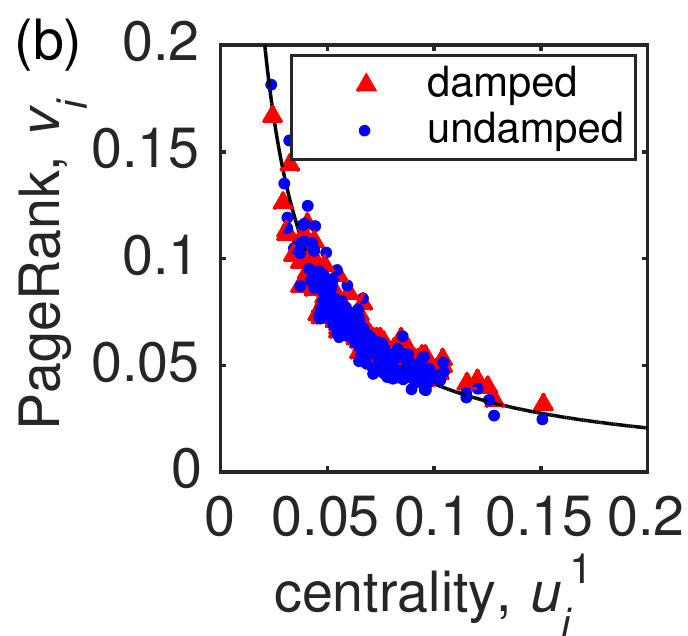}
\caption{(Color online) {\it First-left-singular-vector centrality and PageRank.} (a) Entries $\bm{u}_i^1$ of the first left singular vector vs the out-to-in-degree ratio $k_i^\text{out}/k_i^\text{in}$ for an ER network of size $N=200$ and $p=0.1$. (b) The relationship between PageRank entries $v_i$ (damped and undamped cases are plotted with red triangles and blue dots, respectively) and first-left-singular-vector entries for the same network. The expected inverse relationship $u_i^1v_i\approx\text{const.}$ is plotted as a black curve.} \label{fig3}
\end{figure}

To provide further insight into the structure of $\bm{u}^1$, we consider instead $D_\text{in}^{-1}\tilde{A}^T$, where $\tilde{A}_{ij}=k_i^\text{in}k_j^\text{out}/N\langle k\rangle$ is the mean-field counterpart to $A$. In particular, the corresponding mean-field approximation of $\bm{u}^1$, which satisfies $\tilde{\bm{u}}^1=D_\text{in}^{-1}\tilde{A}^T\tilde{\bm{u}}^1$, is precisely
\begin{align}
\tilde{u}_i^1=ck_i^\text{out}/k_i^\text{in},\label{eq:10}
\end{align}
where $c=[\sum_j(k_j^\text{out}/k_j^\text{in})^2]^{-1/2}$ is a normalizing factor. Thus, the centrality induced by $ {\bm{u}}^1$ can be approximated by the out-to-in-degree ratio $k_i^\text{out}/k_i^\text{in}$ -- a local indicator of the out-flow at a given node. In Fig.~\ref{fig3}(a), we plot the entries $u_i^1$ vs $k_i^\text{out}/k_i^\text{in}$ for an ER network of size $N=200$ with $p=0.2$, and we denote the mean field approximation given by Eq.~(\ref{eq:10}) with a dashed black line. In Fig.~\ref{fig3}(b), we compare the centrality induced by $\bm{u}^1$ to PageRank centrality induced by $\bm{v}$; we plot the entries $v_i$ vs $u_i^1$ for both a damped case ($q=0.15$) and the undamped case ($q=0$) in red triangles and blue dots, respectively. The black curve indicates an approximate inverse relationship between the entries of $\bm v$ and $\bm u^1$. Specifically, we use an approximation similar to the derivation of $\tilde{\bm{u}}^1$ to find $\tilde{\bm{v}}\varpropto k_i^\text{in}/k_i^\text{out}$, which implies that the mean field approximations satisfy
\begin{align}
\tilde{u}_i^1\tilde{v}_i=\left(\sqrt{\sum_{j=1}^N\left(\frac{k_j^\text{out}}{k_j^\text{in}}\right)^2}\sqrt{\sum_{j=1}^N\left(\frac{k_j^\text{in}}{k_j^\text{out}}\right)^2}\right)^{-1},\label{eq:11}
\end{align}
where the right-hand side is a constant. The strong agreement between Eq.~(\ref{eq:11}) and the actual entries of $\bm{u}^1$ and $\bm{v}$ illustrates the strong and opposite relationship between the centrality induced by the first left singular vector $\bm u^1$ and PageRank $\bm v$.

This relationship between synchronization and PageRank that is revealed by the collective frequency variation of a network represents a new direction for network analysis and, immediately, the potential for applying PageRank-based techniques to self-organizing networks. PageRank and random walker dynamics remains one of the most popular topics of research connecting various disciplines, and has a rich literature~\cite{Langville2006}. Specifically, various algorithms and techniques exist for analysis and optimization which might be applied to manipulate a network's collective frequency variation. For instance, given the inverse relationship between the first left singular vector centrality and PageRank, we expect that increasing (decreasing) a node's PageRank corresponds to decreasing (increasing) its contribution to the collective frequency variation.

\section{Power grid dynamics}\label{sec5}

The power grid represents a prime example of a network of self-organizing dynamical systems whose functionality we rely on everyday -- without robust synchronization near a specified range ($\sim$ 50-60 Hz), our power supply is jeopardized. Power grids~\cite{Nishikawa2015NJP} are widely modeled using the following system of second-order differential equations:
\begin{align}
H_i\ddot{\theta}_i+C_i\dot{\theta}_i=P_i+K\sum_{j=1}^NA_{ij}\sin(\theta_j-\theta_i-\alpha_{ij}),\label{eq:12}
\end{align}
where $\theta_i$ represents the mechanical phase of oscillator $i$, $H_i$ and $C_i$ represent the inertial and damping constants, respectively, $P_i$ represents the generated or consumed power of oscillator $i$, $K$ is the global coupling strength, and $\alpha_{ij}$ is a phase-lag parameter for the interaction between oscillators $i$ and $j$. Although the adjacency matrix $A$ is taken to be undirected, the presence of heterogeneity in the damping coefficients yields an effective directedness in the network coupling. Specifically, dividing Eq.~(\ref{eq:12}) through by $C_i$ and linearizing around the synchronized state $\dot{\theta}_1=\dots=\dot{\theta}_N$, where we expect $\theta_1\approx\dots\approx\theta_N$, yields the system
\begin{align}
\tilde{H}_i\ddot{\theta}_i+\dot{\theta}_i=\tilde{\omega}_i-K\sum_{j=1}^N\tilde{L}_{ij}\theta_j,\label{eq:13}
\end{align}
where the new Laplacian $\tilde{L}$ is defined $\tilde{L}_{ij}=\delta_{ij}\tilde{k}_i^{\text{in}}-\tilde{A}$, where $\tilde{A}_{ij}=A_{ij}\cos\alpha_{ij}/C_i$, $\tilde{k}_i^{\text{in}}=\sum_{j}\tilde{A}_{ij}$, $\tilde{H}_i=H_i/C_i$, and $\tilde{\omega}_i=(P_i-K\sum_{j}A_{ij}\sin\alpha_{ij})/C_i$. Note in particular that the effective coupling  matrices are directed, i.e., $\tilde{L}^T\ne\tilde{L}$ and $\tilde{A}^T\ne\tilde{A}$. Different power grid models treat the inertial term in Eq.~(\ref{eq:12}) differently~\cite{Nishikawa2015NJP}. In certain models the inertial term $H_i$ depends on the role of oscillator $i$: if oscillator $i$ is a source, or power generator, $H_i$ is nonzero, but if it is a load, or power consumer, $H_i$ is zero and thus the equation for oscillator $i$ is a first-order differential equation. Some models treat all $H_i$'s as non-zero, resulting in a full system of second-order differential equations, and others treat all $H_i$'s as zero, resulting in a full system of first-order differential equations. We note that regardless of the treatment of the inertial terms, in the synchronized state $\ddot{\theta}_i$ holds for all $i$, and therefore the collective frequency of the synchronized state is preserved, and thus Eq.~(\ref{eq:07}) holds.

\begin{figure*}[t]
\centering
\includegraphics[width=0.95\linewidth]{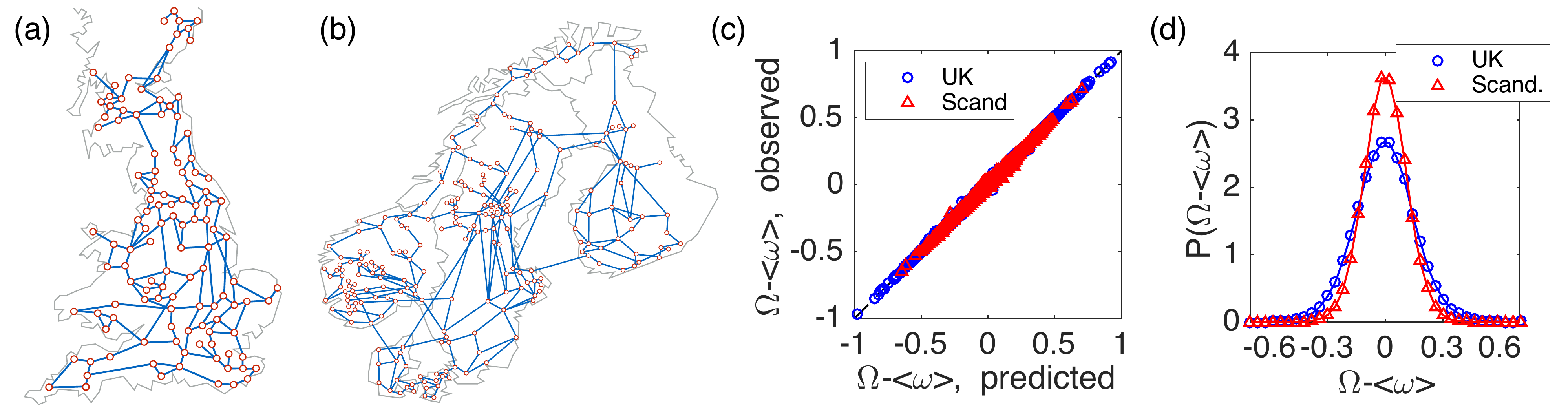}
\caption{(Color online) {\it Collective frequency variation in power grid networks.} (a),(b) Course-grain representations of the UK and Scandinavian power grids, respectively. (c) Collective frequency variation $\Omega-\langle\omega\rangle$ as observed from direct simulations of the power grid model given by Eq.~(\ref{eq:12}) on the UK and Scandinavian power grid networks compared to the theoretical prediction of Eq.~(\ref{eq:07}) given Eq.~(\ref{eq:13}) in $50,000$ realizations. Parameters $P_i$ and $C_i$ are drawn from a bimodal normal distribution and a gamma distribution with mean one, respectively, as described in the text. (d) Distribution of collective frequency variation found for each network.}
\label{fig4}
\end{figure*}

To demonstrate the presence of collective frequency variation in a real-world setting, we consider the power grid model in Eq.~(\ref{eq:12}) on empirical power grids. Specifically, we consider course-grain versions of the UK and Scandinavian power grids~\cite{Rohden2012PRL,Simonsen2008PRL,Witthaut2015Arxiv}, which we illustrate in Fig.~\ref{fig4}(a) and (b), respectively, and which consist of $N=119$ and $236$ nodes and $M=165$ and $320$ links, respectively. It is well-known that in real-world power grids the power $P_i$  of sources and loads are positive and negative with respect to their mean, and damping coefficients are all positive, and with an appropriate rescaling of time can be set to have mean one~\cite{Lozano2012EPJB}. Therefore, we draw each $P_i$ from the bimodal normal distribution $h(P)=(e^{-(P-P_0)^2/2\sigma^2}+e^{-(P+P_0)^2/2\sigma^2})/2\sqrt{2\pi\sigma^2}$ and each $C_i$ from the gamma distribution $g(C)=\alpha^\alpha C^{\alpha-1}e^{-\alpha C}/\Gamma(\alpha)$. For simplicity, inertial coefficients $H_i$ and phase lags $\alpha_{ij}$ are all set to zero. We simulate Eq.~(\ref{eq:12}) using $K=3$, $P_0=\sqrt{3}$, $\sigma=1/2$, and $\alpha=4$ on both the UK and Scandinavian power grid networks, calculating the collective frequency variation $\Omega-\langle\omega\rangle$ from direct observation, and compare to the theoretical prediction of Eq.~(\ref{eq:07}) given Eq.~(\ref{eq:13}) in Fig.~\ref{fig4}(c) for $50,000$ realizations of the parameters $P_i$ and $C_i$. The dashed black curve (which is almost completely covered) underscores perfect agreement. In Fig.~\ref{fig4}(d) we plot the distribution of collective frequencies found in the $50,000$ trials on each network, demonstrating that collective frequency variation can be a significant effect in important, real-world networks such as power grids. Furthermore, our numerical exploration indicate that by appropriately adding and/or deleting links, the collective frequency variation can be either amplified or mitigated, suggesting that the collective frequency variation could be tuned with a collection of judiciously chosen perturbations to the network structure.

\section{Discussion}\label{sec6}

In this Article, we have studied the collective frequency of self-organizing systems in general directed networks. In particular, we have shown that in generic directed networks the collective frequency variation is nonzero and is given by a weighted average of the natural frequencies. In other words, the collective frequency of the synchronized state is not equal to the mean of the oscillators' individual natural frequencies. The weights that determine the collective frequency variation are associated with the left singular vector $\bm{u}^1$ of the Laplacian matrix $L$ corresponding to the singular value $\sigma^1=0$. This formalism allows us to define and calculate the full range of collective frequency variations possible for any given network. We have shown that the only networks with generically zero collective frequency variation are degree-balanced networks in which the in- and out-degrees match for every node (i.e., $k_i^\text{in}=k_i^\text{out}$).

We have found that the first left singular vector in fact induces a centrality measure on the network. This centrality is intimately linked with the directedness of the network and measures an effective out-flow at each node. Interestingly, we have found that this centrality is a reverse analogue of PageRank centrality~\cite{Brin1998}; PageRank is a cornerstone to Google's ranking of webpages and is well-known to quantify the in-flow at each node~\cite{Gleich2015SIAM}. Moreover, we have shown that the mean field approximations to the first-left-singular-vector centrality and the PageRank centrality are precisely the inverse of one another.

We believe that these results will have significant impact on the study of self-organizing processes on networks, since in many application the collective dynamics of the synchronized state, i.e., the collective frequency, plays an important role in the functionality of the system. As a prime example we have considered the dynamics of two real-world power grids -- a particularly important complex network of oscillators (i.e., sources and loads) that governs the flow of energy~\cite{Dorfler2012SIAM}. In particular, power grids must synchronize to avoid power failures, but must also evolve close enough to a reference frequency of approximately $50$ - $60$ Hz~\cite{Lozano2012EPJB}. We have demonstrated that, despite the fact that power grid networks are structurally undirected, dynamical heterogeneity yields an effectively directed network structure, and therefore allows significant collective frequency variation. However, we emphasize that our results have broader applications than just power grid dynamics. In fact, the collective frequency of an ensemble plays a crucial roll in the functionality of a wide range of systems from disciplines including biology, neuroscience, and engineering. Examples of systems whose functionality can be compromised if the collective frequency differs too much from a given reference include oscillations of brain waves, propagation of activity through cardiac tissue, consensus in sensor networks, and the coordination of muscle contractions in the digestive track~\cite{Schnitzler2005NRN,Karma2013,OlaftiSaber007IEEE,Aliev200JTB}.

Moreover, these results demonstrate a novel relationship between a widely used topological quantity used to rank the importance of nodes and the dynamical process of synchronization. The implications point towards a new method of ranking nodes using synchronization -- a notion consistent with other findings where synchronization can be utilized to uncover topological properties of networks~\cite{Arenas2006PRL}. A particularly interesting finding is the link between the synchronization dynamics of a network ensemble and the role of PageRank in determining each oscillator's contribution to the collective frequency. This link opens the possibility for analysis and optimization of the synchronization properties of networks using PageRank -- a topic with a large body of literature and well-established algorithms for optimization~\cite{Langville2006}. In particular, we expect that pre-existing methods for optimizing the PageRank in networks can be applied to manipulate the collective frequency of generic various oscillator networks.

\acknowledgements
We thank M. Timme for the UK and Scandinavian power grid networks. DT acknowledges support from NIH Award Number R01HD075712. JS acknowledges funding from the Simons Foundation Grant No. 318812 and the Army Research Office Grant No. W911NF-12-0276. AA acknowledges support by the European Commission FET-Proactive project MULTIPLEX (Grant No. 317532), the ICREA Academia, the James S. McDonnell Foundation, and by FIS2015-71582-C2-1.

\bibliographystyle{plain}

\end{document}